\documentclass[11pt]{article}
\usepackage{amssymb,amsmath,epsfig}
\def\one{{{{\rm 1} \kern -.19em {\rm l}}}}
\def\C{{{{\rm {\mbox{\small l}}} \kern -.50em {\rm C}}}}
\def\R{{{{\rm l} \kern -.15em {\rm R}}}}
\def\N{{{{\rm l} \kern -.15em {\rm N}}}}
\def\E{{{{\rm l} \kern -.15em {\rm E}}}}
\def\P{{{{\rm l} \kern -.15em {\rm P}}}}
\def\Z{{{{\rm Z} \kern -.35em {\rm Z}}}}
\def\1{{{{\rm 1} \kern -.35em {\rm 1}}}}

\begin{document}
\begin{sloppypar}
\vspace*{0cm}
\begin{center}
{\setlength{\baselineskip}{1.0cm}{ {\Large{\bf
QUANTUM MODELS WITH ENERGY-DEPENDENT POTENTIALS SOLVABLE IN TERMS OF \\ EXCEPTIONAL 
ORTHOGONAL POLYNOMIALS
\\}} }}
\vspace*{1.0cm}
{\large{\sc{Axel Schulze-Halberg}}}$^\dagger$ and {\large{\sc{Pinaki Roy}}}$^\ddagger$
\end{center}
\noindent \\

$\dagger$ Department of Mathematics and Actuarial Science and Department of Physics, Indiana University Northwest, 3400 Broadway,
Gary IN 46408, USA, e-mail: axgeschu@iun.edu, xbataxel@gmail.com \\ \\

$\ddagger$ Physics and Applied Mathematics Unit, Indian Statistical Institute, Kolkata 700108, India, 
e-mail: pinaki@isical.ac.in \\ \\

\vspace*{.5cm}
\begin{abstract}
\noindent
We construct energy-dependent potentials for which the Schr\"odinger equations admit solutions in terms of 
exceptional orthogonal polynomials. Our method of construction is based on certain point transformations, applied to 
the equations of exceptional Hermite, Jacobi and Laguerre polynomials. We present several examples of 
boundary-value problems with energy-dependent potentials that admit a discrete spectrum and the corresponding 
normalizable solutions in closed form.

\end{abstract}


\section{Introduction}
Systems that feature energy-dependent potentials occur frequently in various areas of Quantum Mechanics 
and its applications. Recent examples for such systems can be found in magneto-hydrodynamic models of the dynamo
effect \cite{dynamo}, hydrodynamics \cite{hydro1}, the Hamiltonian formulation of relativistic
quantum mechanics \cite{rela2}, the area of quantum wells and semiconductors
\cite{semi1} \cite{semi2}, and models of heavy quark systems \cite{heavy}, just to name a few. Due to their practical 
applicability, mathematical properties of energy-dependent potentials have been studied thoroughly, especially in regards to the modified quantum 
theory that they obey. In particular, the construction of the completeness relation and the usual $L^2$-norm are 
affected \cite{formanek}. Further theoretical work on energy-dependent potentials includes the application to 
confined models \cite{lombard}, their generation by means of the supersymmetry formalism \cite{yekken} and 
through point transformations of hypergeometric equations \cite{jesus}. Similar to the work done in the latter two 
references, the purpose of this research is to find quantum models featuring energy-dependent potentials. However, in difference to 
former work, here we are interested in models that admit solutions in terms of exceptional orthogonal polynomial systems 
(X-OPS). After the first introduction of these systems \cite{ullate1} \cite{ullate2} as a generalization of conventional 
orthogonal polynomial families, a vast amount of research has been devoted to studying X-OPS. Besides 
being interesting in themselves, they have been used to construct rational extensions of quantum-mechanical potentials 
through the Darboux transformation. 
Some recent applications include the relationship between X-OPS, the Darboux 
transformation and shape invariant potentials \cite{Quesne2} \cite{Odake1} \cite{sree}, recurrence relations for the 
coefficients of exceptional orthogonal polynomials \cite{odarec1}-\cite{odarec3} and quantum superintegrability 
\cite{marquette}. Let us point out that the aforementioned references are exemplary and by far not exhausting. For an 
introduction to the topic, the reader is referred to the work \cite{ullategen}. Since it is of high interest to find 
quantum models that allow for closed-form solutions in terms of X-OPS, in the present note we construct such models, 
focusing particularly on energy-dependent potentials. To this end, we apply a class of point transformations to 
differential equations that are solved by X-OPS, converting the equations into Schr\"odinger form for energy-dependent 
potentials. For the sake of completeness, in section 2 we give a short introduction to X-OPS and we review 
point transformations for linear differential equations of second order. Section 3 is devoted to the actual 
construction of Schr\"odinger-type equations that admit solutions in terms of X-OPS, namely the exceptional Hermite, 
Jacobi and Laguerre polynomials \cite{abram}. We will present several 
examples, in which the underlying Schr\"odinger equation is equipped with boundary conditions, such that the 
resulting boundary-value problem has bound-state solutions and a discrete energy spectrum. The validity of the 
solutions obtained in the examples is verified by calculating the norm for energy-dependent quantum system. In the final 
example we show that in particular cases our method can be adapted to be compatible with the Dirac equation.

\section{Preliminaries}
We will now summarize basic properties of the three X-OPS that we focus on in this work. In addition, we briefly review 
point transformations between second-order equations.

\subsection{Exceptional orthogonal polynomials}
A polynomial family that is orthogonal and complete with respect to the inner product of a weighted $L^2$-Hilbert space 
is called an X-OPS, if the degree sequence of its elements does not contain all 
nonnegative integers. Here, we distinguish the Hermite, Jacobi, and Laguerre systems. For a more detailed discussion 
of these three systems the reader is referred to \cite{ullateh} \cite{midya} \cite{liaw}, respectively. In the following we will 
summarize partial results taken from the aforementioned references.

\paragraph{\boldmath{$X_\lambda$}-Hermite polynomials.} Let $\lambda=(l_j)$ be a sequence of 
$m \geq 1$ pairwise different, nonnegative integers and define $W_\lambda$ as the Wronskian of the Hermite polynomials 
with degrees 
$l_1,l_2,...,l_m$. Then, the X-OPS of $X_\lambda$-Hermite polynomials consists of the elements
\begin{eqnarray}
{\cal H}_{\lambda,n}(x) &=& W_{(H_{l_j})_{j=1}^m,H_n}(x), ~~~n \notin \lambda. \label{xher}
\end{eqnarray}
Here, the right side represents the Wronskian of the Hermite polynomials with degrees $l_1,l_2,...,l_m$ and $n$, 
where $n$ runs through all nonnegative integers except those contained in the sequence $\lambda$. Depending on the 
choice of $\lambda$, the degrees of the $X_\lambda$-Hermite polynomials (\ref{xher}) skip at least one 
nonnegative integer. In addition, they satisfy the equation
\begin{eqnarray}
{\cal H}_{\lambda,n}''(x)-2 \left[
x+\frac{W_\lambda'(x)}{W_\lambda(x)}\right] {\cal H}_{\lambda,n}'(x)+
\left[\frac{W_\lambda''(x)}{W_\lambda(x)}+2~x~\frac{W_\lambda'(x)}{W_\lambda(x)}-2~(m-n)
\right] {\cal H}_{\lambda,n}(x) &=& 0. \label{odeh}
\end{eqnarray}
Furthermore, the $X_\lambda$-Hermite polynomials satisfy
\begin{eqnarray}
\int\limits_{-\infty}^\infty {\cal H}_{\lambda,n_1}^\ast(x)~{\cal H}_{\lambda,n_2}(x)~\frac{\exp\left(-x^2 \right)}{W_\lambda(x)^2}~dx 
~=~ \delta_{n_1 n_2}~C_H, \label{ogh}
\end{eqnarray}
where $C_H$ is a constant and the asterisk denotes complex conjugation. Relation (\ref{ogh}) states that 
the $X_\lambda$-Hermite polynomials form an orthogonal set in the weighted Hilbert space 
$L^2_w(-\infty,\infty)$, where the weight function is given by $w(x)=\exp(-x^2)/W_\lambda(x)^2$.

\paragraph{\boldmath{$X_m$}-Jacobi polynomials.} For a natural number $m$ and two real numbers $a$, $b$, we define 
the X-OPS of $X_m$-Jacobi polynomials as
\begin{eqnarray}
{\cal P}_{m,n}^{(a,b)}(x) ~=~ \frac{(-1)^m}{2~(1+a-m+n)} 
\Bigg[
(1+a+b-m+n)~(x-1)~P_m^{(-a-1,b-1)}(x)~P_{n-m-1}^{(a+2,b)}(x)+ \nonumber \\[1ex]
& & \hspace{-12.5cm} +~2~(1+a-m)~P_m^{(-a-2,b)}(x)~P_{n-m}^{(a+1,b-1)}(x)
\Bigg],~~~n~=~m+1,m+2,..., \label{xjac}
\end{eqnarray}
where the symbol $P_j^{(k,l)}$ for an integer $j$ and real numbers $k,l$ represents a Jacobi polynomial. Inspection of (\ref{xjac}) shows that the degree of the 
$X_m$-Jacobi polynomials is at least $m$, such that all degrees between zero and $m-1$ are skipped. 
For a fixed value of $n$, the associated $X_m$-Jacobi polynomial is a solution of the equation
\begin{eqnarray}
(1-x^2)~{\cal P}_{m,n}^{(a,b)}{''}(x)+ & & \nonumber \\[1ex]
& & \hspace{-4.1cm}+ \Bigg[
(b+1) (1-x)-(a+1) (x+1)+\frac{(1+a-b-m)~(1-x^2)~P_{m-1}^{(-a,b)}(x)}{P_{m}^{(-a-1,b-1)}(x)}
\Bigg]~{\cal P}_{m,n}^{(a,b)}{'}(x)
+ \nonumber \\[1ex]
& & \hspace{-4.1cm}+ \Bigg[
-2~b~m+(1+a+b-2~m)~n+n^2+\frac{b~(a+a-b-m) (1-x)~P_{m-1}^{(-a,b)}(x)}{P_{m}^{(-a-1,b-1)}(x)}
\Bigg]~{\cal P}_{m,n}^{(a,b)}(x)  =  0.  \nonumber \\ \label{odej}
\end{eqnarray}
These polynomials satisfy the orthogonality relation
\begin{eqnarray}
\int\limits_{-1}^1 {\cal P}_{m,n_1}^{(a,b)\ast}(x) ~{\cal P}_{m,n_2}^{(a,b)}(x) ~\frac{(1-x)^a (1+x)^b}{P_m^{(-a-1,b-1)}(x)^2}~dx 
&=& \delta_{n_1 n_2}~C_J, \label{ogj}
\end{eqnarray}
where $C_J$ is a constant. Hence, the $X_m$-Jacobi polynomials form an orthogonal set in the weighted Hilbert space 
$L^2_w(-1,1)$ for the weight function $w(x)=(1-x)^a (1+x)^b/P_m^{(-a-1,b-1)}(x)^2$.

\paragraph{\boldmath{$X_m$}-Laguerre polynomials.} Let $m$ and $\alpha$ be a natural and a real number, 
respectively. We distinguish three types of X-OPS, elements of which are called $X_m$-Laguerre polynomials. 
\begin{itemize}
\item For $\alpha>0$, the type I $X_m$-Laguerre polynomials are defined by
\begin{eqnarray}
{}_1{\cal L}_{m,n}^\alpha(x) ~=~ L_m^\alpha(-x)~L_{n-m}^{\alpha-1}(x)+L_{m}^{\alpha-1}(-x)~L_{n-m-1}^{\alpha}(x), 
~~~n=m,m+1,m+2,...,
\label{xlag1}
\end{eqnarray}
where $L_j^k$ for an integer $j$ and a real number $k$ represents an associated Laguerre polynomial. We observe that the degree of ${}_1{\cal L}_{m,n}^\alpha$ is at least $m$, since 
$n$ must be greater or equal to $m$. The polynomials in (\ref{xlag1}) solve the differential equation
\begin{eqnarray}
-x~{}_1{\cal L}_{m,n}^\alpha{''}(x)+\Bigg[
x-\alpha-1+\frac{2~x~L_m^{\alpha-1}{'}(-x)}{L_m^{\alpha-1}(-x)}
\Bigg]~{}_1{\cal L}_{m,n}^\alpha{'}(x)+ & & \nonumber \\[1ex]
& & \hspace{-4cm} +~
\Bigg[
-n+\frac{2~\alpha~L_m^{\alpha-1}{'}(-x)}{L_m^{\alpha-1}(-x)}
\Bigg]~{}_1{\cal L}_{m,n}^\alpha(x) ~=~ 0. \label{odel1}
\end{eqnarray}
Furthermore, they satisfy the orthogonality relation
\begin{eqnarray}
\int\limits_0^\infty {}_1{\cal L}_{m,n_1}^{\alpha~ \ast}(x)~{}_1{\cal L}_{m,n_2}^\alpha(x)~\frac{x^\alpha~ \exp(-x)}
{L_m^{\alpha-1}(-x)^2}~dx &=& \delta_{n_1 n_2}~C_{L_1}, \nonumber
\end{eqnarray}
for a constant $C_{L_1}$. Thus, they form an orthogonal 
set in the weighted Hilbert space $L^2_w(0,\infty)$, where the weight function is given by 
$w(x)=x^\alpha \exp(-x)/L_m^{\alpha-1}(-x)^2$.

\item For $\alpha>m-1$, the type II $X_m$-Laguerre polynomials are defined by
\begin{eqnarray}
{}_2{\cal L}_{m,n}^\alpha(x) ~=~ x~L_m^{-\alpha-1}(x)~L_{n-m-1}^{\alpha+2}(x)+
(m-\alpha-1)~L_{m}^{-\alpha-2}(x)~L_{n-m}^{\alpha+1}(x), & &\nonumber \\[1ex]
& & \hspace{-3.1cm} n=m,m+1,m+2,...
\label{xlag2}
\end{eqnarray}
Similar to the previous type, we observe that the degree of ${}_2{\cal L}_{m,n}^\alpha$ is at least $m$. 
The polynomials in (\ref{xlag2}) satisfy
\begin{eqnarray}
-x~{}_2{\cal L}_{m,n}^\alpha{''}(x)+\Bigg[
x-\alpha-1+\frac{2~x~L_m^{\alpha-1}{'}(-x)}{L_m^{\alpha-1}(-x)}
\Bigg]~{}_2{\cal L}_{m,n}^\alpha{'}(x)+ & & \nonumber \\[1ex]
& & \hspace{-4cm} +~
\Bigg[
-n+\frac{2~\alpha~L_m^{\alpha-1}{'}(-x)}{L_m^{\alpha-1}(-x)}
\Bigg]~{}_2{\cal L}_{m,n}^\alpha(x) ~=~ 0. \label{odel2}
\end{eqnarray}
Furthermore, they fulfill the orthogonality relation
\begin{eqnarray}
\int\limits_0^\infty {}_2{\cal L}_{m,n_1}^{\alpha~ \ast}(x)~{}_2{\cal L}_{m,n_2}^\alpha(x)~\frac{x^\alpha~ \exp(-x)}
{L_m^{-\alpha-1}(x)^2}~dx &=& \delta_{n_1 n_2}~C_{L_{2}}, \nonumber
\end{eqnarray}
for a constant $C_{L_{2}}$. Thus, they form an orthogonal 
set in the weighted Hilbert space $L^2_w(0,\infty)$, where the weight function is given by 
$w(x)=x^\alpha \exp(-x)/L_m^{-\alpha-1}(x)^2$.

\item For $-1<\alpha<0$, the type III $X_m$-Laguerre polynomials are defined by
\begin{eqnarray}
{}_3{\cal L}_{m,0}^\alpha(x) &=& 1 \nonumber \\[1ex] 
{}_3{\cal L}_{m,n}^\alpha(x) &=& x~L_{n-m-2}^{\alpha+2}(x)~L_{m}^{-\alpha-1}(-x)+
(m+1)~L_{n-m-1}^{\alpha+1}(x)~L_{m+1}^{-\alpha-2}(-x), \nonumber \\[1ex]
& & \hspace{6.5cm} n=m+1,m+2,m+3,...
\label{xlag3}
\end{eqnarray}
Here it is understood that $L_{-1}^\alpha = 0$. We remark that the degree of ${}_3{\cal L}_{m,n}^\alpha$ is at least 
$m+1$ if $n \neq 0$. The polynomials in (\ref{xlag3}) satisfy
\begin{eqnarray}
-x~{}_3{\cal L}_{m,n}^\alpha{''}(x)+\Bigg[
-1-\alpha+x+\frac{2~x~L_m^{-\alpha-1}{'}(-x)}{L_m^{-\alpha-1}(-x)}
\Bigg]~{}_3{\cal L}_{m,n}^\alpha{'}(x)-n~{}_3{\cal L}_{m,n}^\alpha(x) ~=~ 0. \label{odel3}
\end{eqnarray}
Furthermore, they satisfy the orthogonality relation
\begin{eqnarray}
\int\limits_0^\infty {}_3{\cal L}_{m,n_1}^{\alpha~ \ast}(x)~{}_3{\cal L}_{m,n_2}^\alpha(x)~\frac{x^\alpha~ \exp(-x)}
{L_m^{-\alpha-1}(-x)^2}~dx &=& \delta_{n_1 n_2}~C_{L_3}, \nonumber
\end{eqnarray}
for a constant $C_{L_3}$. Thus, they form an orthogonal 
set in the weighted Hilbert space $L^2_w(0,\infty)$, where the weight function is given by 
$w(x)=x^\alpha \exp(-x)/L_m^{-\alpha-1}(-x)^2$.
\end{itemize}

\subsection{Point transformations}
In section 3 we will convert the differential equations for X-OPS into Schr\"odinger form using a 
specific point transformation. For the sake of completeness, we now present the explicit forms of this 
transformation and of the transformed equation. 
We assume that a linear differential equation of second order in the generic form
\begin{eqnarray}
f(x)~\chi''(x)+g(x)~\chi'(x)+h(x)~\chi(x) &=& 0, \label{ode}
\end{eqnarray}
is given, where the prime denotes differentiation. We require the functions $f$ and $g$ to be continuously differentiable on an 
open real interval $I$, while $h$ must be continuous on that interval. Furthermore, 
we require any solution $\chi$ to be twice continuously differentiable on $I$. Next, we introduce a 
function $x=x(y)$ that is one-to-one and three times continuously differentiable on a real interval $J$, and has preimage $I$ 
(coordinate change). Upon defining
\begin{eqnarray}
\Phi(y) &=&  \exp\left\{ \int\limits^y
\frac{g\left[x(t)\right] x'(t)}{2~f\left[x(t)\right]}~dt\right\}~
\sqrt{\frac{1}{x'(y)}}~
\chi\left[x(y)\right], \label{pct}
\end{eqnarray}
the function $\Phi$ solves the following differential equation on the interval $J$
\begin{eqnarray}
\Phi''(y) +\Bigg\{
-\frac{g\left[x(y)\right]^2 \left[ x'(y) \right]^2}{4~f\left[x(y)\right]^2}+\frac{h\left[x(y)\right]~\left[x'(y) \right]^2}{f\left[x(y)\right]}+
\frac{g\left[x(y)\right]~f'\left[x(y)\right] \left[ x'(y) \right]^2}{2~f\left[x(y)\right]^2}- \nonumber \\[1ex]
& & \hspace{-9.5cm} -~\frac{g'\left[x(y)\right] \left[x'(y)\right]^2}{2~f\left[x(y)\right]}-\frac{3 \left[x''(y) \right]^2}{4~\left[x'(y) \right]^2}+
\frac{x'''(y)}{2~x'(y)}
\Bigg\} ~\Phi(y) ~=~ 0. \label{odet}
\end{eqnarray}
Note that here the coefficient of $\Phi$ is assumed to be defined on the whole interval $J$.

\section{Quantum models solvable through X-OPS}
We are now ready to construct Schr\"odinger equations that admit solutions in terms of X-OPS. To this end, we will 
first apply our point transformation in its general form (\ref{pct}) to each equation solved by an X-OPS. 
In a subsequent step, we must specify the coordinate change contained in the point transformation, such that the 
resulting Schr\"odinger equation allows for a discrete spectrum and for associated solutions of bound-state type.

\subsection{\boldmath{$X_\lambda$}-Hermite polynomials}
Our first application concerns the differential equation (\ref{odeh}). We will show its general form after undergoing 
the transformation (\ref{pct}), then present a particular example.

\subsubsection{Schr\"odinger-type equation and solutions}
It is straightforward to verify that (\ref{odeh}) is a particular case of the general equation (\ref{ode}) if the following 
settings are used:
\begin{eqnarray}
f(x) ~=~ 1 \qquad g(x) ~=~ -2 \left[
x+\frac{W_\lambda'(x)}{W_\lambda(x)}\right]
\qquad
h(x) ~=~ \frac{W_\lambda''(x)}{W_\lambda(x)}+2~x~\frac{W_\lambda'(x)}{W_\lambda(x)}-2~(m-n). \label{fh}
\end{eqnarray}
The most general Schr\"odinger equation that admits solutions in terms of $X_\lambda$-Hermite polynomials and 
at the same time is related to its counterpart (\ref{odeh}) by means of a point transformation, can be 
found by plugging the coefficients (\ref{fh}) into 
(\ref{odet}). Leaving the coordinate change arbitrary, we obtain
\begin{eqnarray}
\Phi''(y) + \Bigg\{
\left[x'(y)\right]^2 \left[1-2~m+2~n-x(y)^2\right]
-\frac{2~W_\lambda'\left[x(y)\right]^2\left[x'(y)\right]^2}{W_\lambda\left[x(y)\right]^2}+ & & \nonumber \\[1ex]
& & \hspace{-8cm} +~
\frac{2~W_\lambda''\left[x(y)\right] \left[x'(y)\right]^2}{W_\lambda\left[x(y)\right]}
-\frac{3 \left[x''(y) \right]^2}{4~\left[x'(y) \right]^2}+
\frac{x'''(y)}{2~x'(y)}
\Bigg\}~\Phi(y) ~=~ 0. \label{odehs}
\end{eqnarray}
While this equation has Schr\"odinger form, the coefficient of $\Phi$ is not yet guaranteed to feature a constant 
term that can be related to the energy of the underlying system. Such a term must be generated by an appropriate 
choice of the coordinate change $x=x(y)$. Although a rigorous analysis of all possible choices is not feasible, 
inspection of (\ref{odehs}) shows that a linear or exponential coordinate change will in general create a constant 
term. Before we confirm this by presenting an example, let us first comment on the solutions of (\ref{odehs}). They 
are given by the formula (\ref{pct}) after substitution of the settings in (\ref{fh}) and taking into 
account that the function $\chi$ is an $X_\lambda$-Hermite polynomial. We get
\begin{eqnarray}
\Phi(y) &=& \exp\left[-\frac{1}{2}~x(y)^2 \right] \frac{1}{W_\lambda\left[x(y)\right]}~\sqrt{\frac{1}{x'(y)}}~
{\cal H}_{\lambda,n}\left[x(y)\right]. \label{solh}
\end{eqnarray}
We remark that for now this solution is of formal nature, since we are unable to specify its domain unless further 
information is known about the coordinate change $x=x(y)$.

\subsubsection{Example}
\begin{figure}[h]
\begin{center}
\epsfig{file=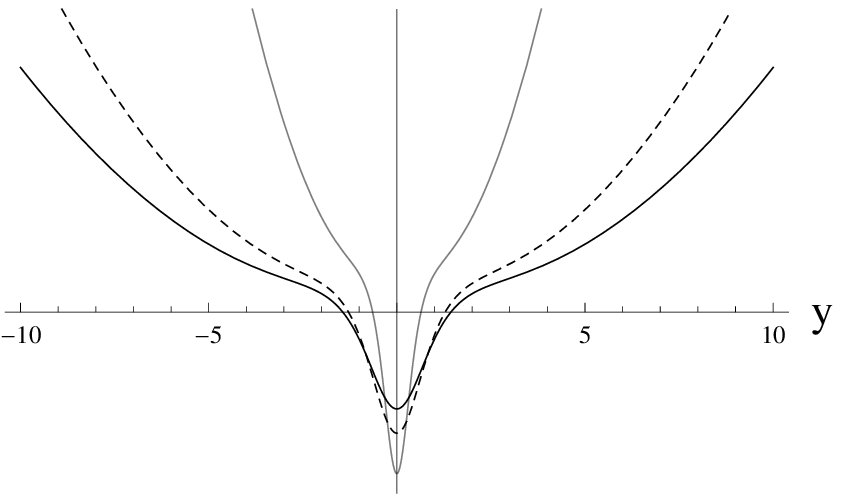,width=7.8cm}
\epsfig{file=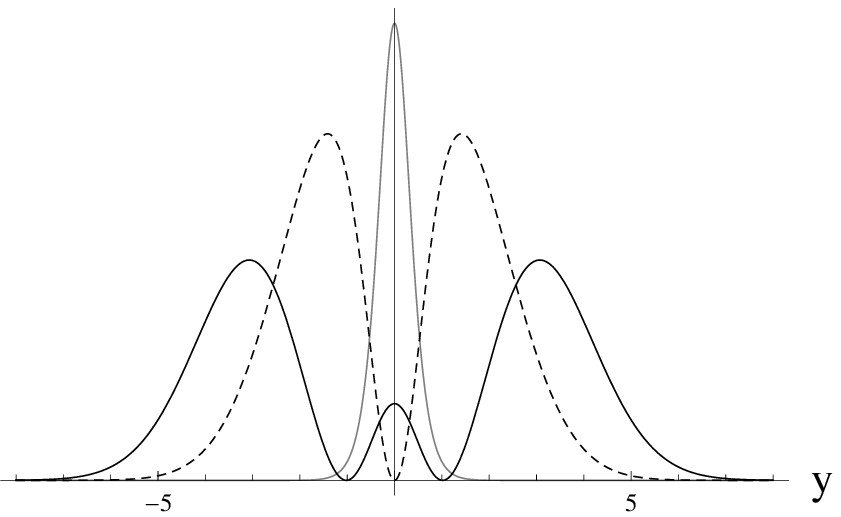,width=7.8cm}
\caption{Vertically scaled graphs of the potential (\ref{poth}) (left plot) and the conventional density $|\Phi_n|^2$ from 
(\ref{solh}) (right plot) for the parameters $n=0$ (grey curve), $n=3$ (dashed curve) and $n=4$ (black curve).}
\label{figh}
\end{center}
\end{figure}

We will now evaluate equation (\ref{odehs}) and its solutions (\ref{solh}) for a particular choice of parameters. Let us 
set
\begin{eqnarray}
\lambda~=~\{1,2\} \qquad \qquad x(y) ~=~ \sqrt{\frac{1}{n+1}}~y. \label{seth}
\end{eqnarray}
We chose a partition of two elements in order to avoid complicated calculations resulting from a large determinant 
$W_\lambda$. Furthermore, the linear coordinate change in (\ref{seth}) guarantees existence of a constant term in 
the potential of our system, which can be related to the stationary energies. 
We observe that for fixed value of $n$ the coordinate change maps the real axis onto itself. 
Upon substitution into (\ref{odehs}) we obtain
\begin{eqnarray}
\Phi''(y)+ \Bigg[
2-\frac{5}{1+n}-\frac{y^2}{(n+1)^2}+\frac{8}{2~y^2+n+1}-\frac{32~y^2}{(2~y^2+n+1)^2}
\Bigg]~\Phi(y) ~=~0. \label{odehsx}
\end{eqnarray}
This result can be interpreted as a Schr\"odinger equation for an energy-dependent potential. Since the domain of our 
equation is the real line, we impose boundary conditions of Dirichlet type at the infinities, that is, we require
\begin{eqnarray}
\lim\limits_{y \rightarrow -\infty} \Phi(y) ~=~\lim\limits_{y \rightarrow \infty} \Phi(y) ~=~0. \label{bvph}
\end{eqnarray}
We see from (\ref{odehsx}) that a constant term was generated in the coefficient of $\Phi$. This term will now play 
the role of the system's energy $E_n$. 
We identify
\begin{eqnarray}
E_n &=& 2-\frac{5}{1+n},~~~n=0,3,4,5,... \label{eneh}
\end{eqnarray}
Here we have included an index on the left side in order to emphasize that the energy is parametrized by the number $n$, 
which can attain all nonnegative integer values except for those contained in $\lambda$, see (\ref{xher}). 
We observe that the energies in 
(\ref{eneh}) form an infinite, strictly increasing sequence that is bounded by the ground state $E_0=-3$ and the accumulation point 
$\lim\limits_{n \rightarrow \infty} E_n = 2$. Inspection of (\ref{odehsx}) confirms that the system's potential is energy-dependent, 
as it contains the parameter $n$. In order to obtain the actual form of the potential, we must first express this parameter 
through the energies by inverting (\ref{eneh}). Note that this is possible because the energy $E_n$ 
is strictly increasing with $n$. We get
\begin{eqnarray}
n &=& -\frac{E_n+3}{E_n-2}. \label{ne}
\end{eqnarray}
Substitution of this result into the coefficient of $\Phi$ in (\ref{odehsx}) yields the energy-dependent potential in the form
\begin{eqnarray}
V(y) &=& \frac{(E_n-2)^2}{25}~y^2+\frac{8~(E_n-2)}{2~E_n~y^2-4~y^2-5}+\frac{80~(E_n-2)}{(2~E_n~y^2-4~y^2-5)^2}. \label{poth}
\end{eqnarray}
A few examples of this potential are shown in the left part of figure \ref{figh}. Note that the shape of the potential changes 
along with the energy. For better visibility we applied a vertical scaling to the plot. 
In the next step, let us compute the solutions of our boundary-value problem (\ref{odehsx}), (\ref{bvph}) that are 
associated with the energies (\ref{eneh}). To this end, we 
plug the settings (\ref{seth}) into the general form of the solution (\ref{solh}). This gives
\begin{eqnarray}
\Phi_n(y) &=& \exp\left(-\frac{y^2}{2~n+2}\right)~\frac{1}{2~y^2+n+1}~{\cal H}_{\{1,2\},n}\left(\sqrt{\frac{1}{n+1}}~y \right). 
\label{solhx}
\end{eqnarray}
Note that we included an index $n$ on the left side and that irrelevant overall factors were omitted. 
The explicit representation of this solution can be obtained from the definition (\ref{xher}) of the 
$X_\lambda$-Hermite polynomials. Upon inserting this definition, we find
\begin{eqnarray}
\Phi_n(y) &=& \exp\Big(-\frac{y^2}{2~n+2}\Big)~\frac{1}{2~y^2+n+1}~\Bigg\{
(n-1)~n~(2~y^2+n+1)~H_{n-2}\left(\sqrt{\frac{1}{n+1}}~y \right)- \nonumber \\[1ex] 
&-& (n+1)~\Bigg[
2~n~\sqrt{\frac{1}{n+1}}~y~H_{n-1}\left(\sqrt{\frac{1}{n+1}}~y \right)-H_{n}\left(\sqrt{\frac{1}{n+1}}~y \right)
\Bigg]
\Bigg\}. \label{phisolh}
\end{eqnarray}
The right part of figure \ref{figh} shows these solutions by means of their conventional density $|\Phi_n|^2$ for the first 
few values of the parameter $n$. We observe that the solutions spread further in horizontal direction as the associated 
potential opens wider upward. Before we conclude this example, let us comment on orthogonality and normalizability 
of the functions (\ref{phisolh}). Since the underlying potential is energy-dependent, we must resort to modified 
definitions of the aformentioned concepts \cite{formanek}. Orthogonality of the family $(\Phi_n)$ holds if the condition
\begin{eqnarray}
\int\limits_J \left[1-\frac{V(y)_{\mid n=n_2}-V(y)_{\mid n=n_1}}{E_{n_2}-E_{n_1}}\right] \Phi_{n_2}^\ast(y)~
\Phi_{n_1}(y)~dy &=& k~\delta_{n_2 n_1}, \label{ortho}
\end{eqnarray}
is fulfilled, where $k$ is a positive constant. The left side of (\ref{ortho}) generates an expression for the norm if the 
limit $n_2 \rightarrow n_1$ is taken. After renaming $n_1$ to $n$, we obtain
\begin{eqnarray}
\Vert \Phi_n \Vert &=& \int\limits_{J} \left[1-\frac{\partial V(y)}{\partial E_n} \right] \left| \Phi_n(y) \right|^2~dy. 
\label{norm}
\end{eqnarray}
It is important to point out that $\Vert \cdot \Vert$ is not a norm in the mathematical sense \cite{formanek} 
because it can become negative, dependent on the 
particular form of the potential $V$. However, in order to render the underlying problem physically meaningful, the 
norm (\ref{norm}) must be positive. Returning to the present example, substitution of the functions (\ref{phisolh}) into 
(\ref{ortho}) shows orthogonality of the family $(\Phi_n)$, recall that the domain of our boundary-value problem is 
$J=(-\infty,\infty)$. We will not show the explicit expressions of the integrand in 
(\ref{ortho}) because they are very large. Instead, let us now determine this norm for the present example. To this end, we must 
plug the potential in its form (\ref{poth}) into the right side of (\ref{norm}), along with the solutions (\ref{solhx}). 
Evaluation of the derivative and reinstalling the parameter $n$ through (\ref{eneh}) gives
\begin{eqnarray}
\left[1-\frac{\partial V(y)}{\partial E_n} \right] \left|\Phi_n(y) \right|^2 &=& 
\exp\Big(-\frac{y^2}{2~n+2}\Big)~\frac{(n+1)^\frac{3}{2}}{80~(2~y^2+n+1)^5} ~
\Bigg[
16~y^8+64~y^6~(n+1)+ \nonumber \\[1ex]
& & \hspace{-3cm} +~72~y^4~(n+1)^2+80~y^2~(n+1)^3-3~(n+1)^4
\Bigg] \left| {\cal H}_{\{1,2\},n}\left(\sqrt{\frac{1}{n+1}}~y \right) \right|^2. \nonumber
\end{eqnarray}
As (\ref{norm}) shows, we must now find the integral of this expression. Since we were not able to perform the 
direct integration without specifying a numerical value for $n$, let us show the first few values of the 
energy-dependent norm integral (\ref{norm}). We have 
\begin{eqnarray}
\Vert \Phi_0 \Vert ~=~ \frac{16}{5}~\sqrt{\pi} \qquad \qquad \qquad 
\Vert \Phi_3 \Vert ~=~ \frac{12288}{5}~\sqrt{\pi} \qquad \qquad \qquad \Vert \Phi_4 \Vert ~=~ 92160~\sqrt{\pi}. \label{hnorms}
\end{eqnarray}
Upon using higher values of $n$, the associated norm integrals increase, implying that they remain positive. 
We can therefore conclude that the solutions (\ref{solhx}) of our boundary-value problem (\ref{odehsx}), (\ref{bvph}) 
are physically correct.
 \noindent 
Before we conclude this example, let us determine the norm (\ref{norm}) in an alternative way that is not 
based on integration. The following general relationship was proven in \cite{xbat}:
\begin{eqnarray}
\Vert \Phi_n \Vert &=& W_{\Phi_n,\frac{\partial \Phi_n}{\partial E_n}}(y_0)-W_{\Phi_n,\frac{\partial \Phi_n}{\partial E_n}}(y_1). \label{nrep}
\end{eqnarray}
Here, $W$ represents the Wronskian of the functions in its index. Furthermore, the quantities $y_0$ and $y_1$ 
represent the lower and higher endpoints of the integration interval $J$. The partial derivative is understood by 
means of the chain rule as 
\begin{eqnarray}
\frac{\partial \Phi_n}{\partial E_n} &=& \frac{\partial \Phi_n}{\partial n} \frac{\partial n}{\partial E_n}, \nonumber
\end{eqnarray}
where the function $n=n(E_n)$ is given in (\ref{ne}). Representation 
(\ref{nrep}) has the advantage that no integral needs to be evaluated and that the potential in (\ref{norm}) does not 
need to be taken into account. In the present case, $\Phi_n$ is the function given in (\ref{phisolh}), while its derivative 
with respect to the parameter $E$ can be constructed by first replacing the parameter $n$ by (\ref{ne}), then taking 
the derivative, and in the final step reinstalling $n$ by substitution of (\ref{eneh}). The numbers $y_0$ and $y_1$ 
are negative and positive infinity, respectively. As a consequence, the right side of (\ref{nrep}) is to be understood in the 
sense of a limit. While the latter expression is a symbolic resolution of the norm integral that can not be 
obtained by integration, its form is very long, such that we omit to state it here explicitly. Instead, we evaluate the 
particular case $n=0$. We find for the Wronskian
\begin{eqnarray}
W_{\Phi_n,\frac{\partial \Phi_n}{\partial E}}(y)_{\mid n=0} &=& \frac{16~\exp\left(-y^2 \right) 
\left[y \left(5+4~y^2+4~y^4\right)+4 \left(1+2~y^2\right)^3 H_{-3}(y)\right]}{5 \left(1+2~y^2\right)^4}. \nonumber
\end{eqnarray}
Here, $H_{-3}$ stands for a Hermite function. Note that this function is not a polynomial, but a 
linear combination of hypergeometric functions \cite{abram}. Application of the limits at positive and negative infinity gives
\begin{eqnarray}
\lim\limits_{y \rightarrow -\infty} W_{\Phi_n,\frac{\partial \Phi_n}{\partial E}}(y)_{\mid n=0} ~=~ \frac{16}{5}~\sqrt{\pi} 
\qquad \qquad \qquad
\lim\limits_{y \rightarrow \infty} W_{\Phi_n,\frac{\partial \Phi_n}{\partial E}}(y)_{\mid n=0} ~=~ 0. \nonumber
\end{eqnarray}
Substitution of these limits into (\ref{nrep}) yields
\begin{eqnarray}
\Vert \Phi_0 \Vert &=& \lim\limits_{y \rightarrow -\infty} W_{\Phi,\frac{\partial \Phi}{\partial E}}(y)_{\mid n=0}-
\lim\limits_{y \rightarrow \infty} W_{\Phi,\frac{\partial \Phi}{\partial E}}(y)_{\mid n=0} ~=~ \frac{16}{5}~\sqrt{\pi}, 
\nonumber
\end{eqnarray}
confirming the result found in (\ref{hnorms}).

\subsection{\boldmath{$X_m$}-Jacobi polynomials}
We now turn to equation (\ref{odej}). As in the previous case of the X-Hermite polynomials, we will bring the 
equation to Schr\"odinger form, adjust the existing solutions, and afterwards present a specific example for a 
case that admits bound-states.

\subsubsection{Schr\"odinger-type equation and solutions}
The governing equation (\ref{odej}) of the present example is a particular case of (\ref{ode}), provided we apply the 
following parameter settings
\begin{eqnarray}
f(x) &=& 1-x^2 \label{fj} \\[1ex]
g(x) &=& \frac{(a-b-m+1)~(1-x^2)~P_{m-1}^{(-a,b)}(x)}{P_{m}^{(-a-1,b-1)}(x)}-(a+1)~(x+1)+(b+1)~(1-x) \label{gj} \\[1ex]
h(x) &=& \frac{b~(a-b-m+1)~(1-x)~P_{m-1}^{(-a,b)}(x)}{P_{m}^{(-a-1,b-1)}(x)}+n^2+n~(a+b-2~m+1)-2~b~m. \label{hj}
\end{eqnarray}
If we plug these choices into the general form of a Schr\"odinger equation (\ref{odet}) that was obtained through a 
point transformation, the resulting equation reads
\begin{eqnarray}
\Phi''(y)+\Bigg\{\hspace{-.4cm}
&-& \hspace{-.2cm}\frac{1}{2~\left[1-x(y)^2 \right]}+
\frac{\left[-2~b~m+(1+a+b-2~m)~n+n^2\right] \left[x'(y) \right]^2]}{1-x(y)^2}+  \nonumber \\[1ex]
&+& \hspace{-.2cm}
\frac{b~(1+a-b-m) \left[1-x(y) \right] \left[x'(y) \right]^2 P_{m-1}^{(-a,b)}\left[x(y) \right] }{\left[1-x(y)^2 \right] 
P_m^{(-a-1,b-1)}\left[x(y) \right]}
-(2+a+b)~\left[x'(y) \right]^2- \nonumber \\[1ex]
&-&  \hspace{-.2cm}
\frac{2~(1+a-b-m)~\left[x'(y) \right]^2 x(y)~P_{m-1}^{(-a,b)}\left[x(y) \right]}{P_m^{(-a-1,b-1)}\left[x(y) \right]}-
\nonumber \\[1ex]
&-&  \hspace{-.2cm}
\frac{(1+a-b-m)~(-1-a+b+m)\left[x'(y) \right]^2 \left[1-x(y)^2 \right] P_{m-1}^{(-a,b)}\left[x(y) \right]^2}
{2~P_{m}^{(-a-1,b-1)}\left[x(y) \right]^2}+
\nonumber \\[1ex]
&+&  \hspace{-.2cm}
\frac{(1+a-b-m)~(-a+b+m) \left[x'(y) \right]^2 \left[1-x(y)^2 \right] P_{m-2}^{(1-a,b+1)}\left[x(y) \right]}
{2~P_{m}^{(-a-1,b-1)}\left[x(y) \right]}-
\nonumber \\[1ex]
&-&  \hspace{-.2cm}
\frac{(b+1)~\left[1-x(y) \right] x(y) \left[x'(y) \right]^2-(a+1)~\left[1+x(y) \right] x(y) \left[x'(y) \right]^2}{\left[1-x(y)^2 \right]^2}+
\nonumber \\[1ex]
&+&  \hspace{-.2cm}
\frac{(1+a-b-m)~x(y) \left[1-x(y)^2 \right] \left[x'(y) \right]^2 P_{m-1}^{(-a,b)}\left[x(y) \right]}
{\left[1-x(y)^2 \right]^2 P_m^{(-a-1,b-1)}\left[x(y) \right]^2}-
\nonumber \\[1ex]
&-&  \hspace{-.2cm}
\frac{1}{4}~
\Bigg[
\frac{(b+1)~\left[1-x(y) \right] x(y) \left[x'(y) \right]^2-(a+1)~\left[1+x(y) \right] x(y) \left[x'(y) \right]^2}{\left[1-x(y)^2 \right]}+
\nonumber \\[1ex]
&+&  \hspace{-.2cm}
\frac{(1+a-b-m)~x(y) \left[1-x(y)^2 \right] \left[x'(y) \right]^2 P_{m-1}^{(-a,b)}\left[x(y) \right]}
{\left[1-x(y)^2 \right] P_m^{(-a-1,b-1)}\left[x(y) \right]}
\Bigg]^2-\nonumber \\[1ex]
&-&  \hspace{-.2cm}
\frac{3 \left[x''(y) \right]^2}{4~\left[x'(y) \right]^2}+
\frac{x'''(y)}{2~x'(y)}~
\Bigg\} ~\Phi(y) ~=~ 0. \label{odejx}
\end{eqnarray}
In the next step we must determine a coordinate change $x=x(y)$ that generates a constant term in the 
coefficient of $\Phi$. While a general analysis of admissible coordinate changes is beyond the scope of this work, 
we observe that (\ref{odejx}) contains a term $\sim x'''/x'$. If we pick an exponential function for the coordinate change, 
the latter term will deliver the sought constant. This will be demonstrated in a particular example below. Now, the 
solutions of equation (\ref{odejx}) can be constructed through the point transformation (\ref{pct}). After inserting the 
present settings (\ref{fj})-(\ref{hj}), we get the result
\begin{eqnarray}
\Phi(y) &=& \left[ x(y)-1\right]^{\frac{a+1}{2}} \left[ x(y)+1\right]^{\frac{b+1}{2}} 
\frac{1}{P_m^{(-a-1,b-1)}\left[x(y) \right]}~\sqrt{\frac{1}{x'(y)}}~{\cal P}_{m,n}^{(a,b)}\left[x(y)\right]. \label{solj}
\end{eqnarray}
Let us point out again that normalizability is not guaranteed until our coordinate change is further specified.

\subsubsection{Example}
\begin{figure}[h]
\begin{center}
\epsfig{file=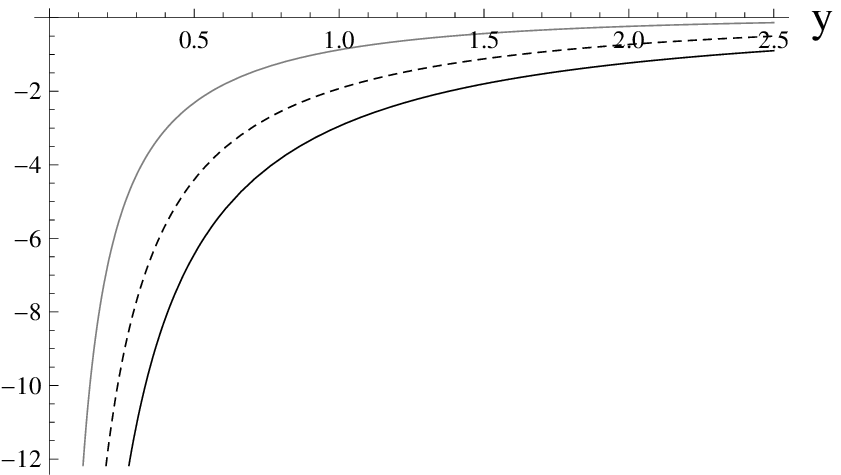,width=7.8cm}
\epsfig{file=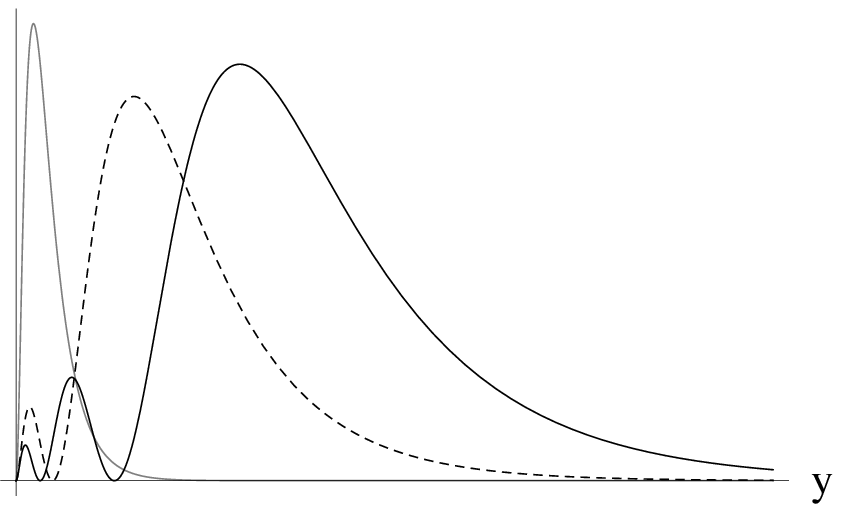,width=7.8cm}
\caption{Graphs of the potential (\ref{potjx}) (left plot) and the conventional density $|\Phi_n|^2$ from 
(\ref{solj}) (right plot) for the parameters $n=2$ (grey curve), $n=4$ (dashed curve) and $n=6$ (black curve). 
For better visibility, the plots were scaled in both horizontal and vertical direction.}
\label{figj}
\end{center}
\end{figure}

In order to work a particular example of the present equation (\ref{odejx}), we must make a choice for the 
parameters and the coordinate change. Let us set
\begin{eqnarray}
a~=~b~=~m~=~1 \qquad \qquad \qquad x(y) ~=~ \exp\left(-\frac{y}{n} \right),~y>0. \label{setj}
\end{eqnarray}
Recall that in the previous paragraph we found an exponential function to be suitable as a coordinate change because 
it will produce a constant energy term in the resulting Schr\"odinger equation. We demonstrate this by plugging 
(\ref{setj}) into the Schr\"odinger-type equation (\ref{odejx}), which gives
\begin{eqnarray}
\Phi''(y) + \Bigg\{
-\frac{1}{4~n^2}+\frac{(n+1)}{n \left[\exp\left(\frac{2~y}{n} \right)-1\right]}
\Bigg\}~\Phi(y) ~=~ 0. \label{odejxx}
\end{eqnarray}
This equation has Schr\"odinger form and features an energy-dependent potential. Observe that 
the coordinate change in (\ref{setj}) maps the positive real axis onto the positive real axis because we restricted the 
exponential function's domain. We can therefore endow our governing equation (\ref{odejxx}) with Dirichlet-type 
boundary conditions
\begin{eqnarray}
\lim\limits_{y \rightarrow 0} \Phi(y) ~=~ \lim\limits_{y \rightarrow \infty} \Phi(y) ~=~ 0. \label{bvpj}
\end{eqnarray}
The anticipated constant term in our equation (\ref{odejxx}) is now identified with the stationary energies of the system. 
In order to emphasize the dependence 
of these energies on the parameter $n$, we include it as an index, that is, we define
\begin{eqnarray}
E_n &=& -\frac{1}{4~n^2}, \label{enej}
\end{eqnarray}
where the parameter $n$ must be at least equal to $m+1$, which equals 2 under the present settings (\ref{setj}). 
We observe that the energies form an infinite sequence bounded from below by the ground state energy $E_2=-1/16$. 
Since the potential in equation (\ref{odejxx}) is not expressed through the actual energy, but the parameter $n$, we 
will rewrite the potential by expressing $n$ through $E_n$. We obtain
\begin{eqnarray}
n &=& \sqrt{-\frac{1}{4~E_n}}. \label{nex}
\end{eqnarray}
Note that the right side is real-valued because $E_n$ is negative. Substitution of (\ref{nex}) into (\ref{odejxx}) and 
extraction of the potential gives
\begin{eqnarray}
V(y) &=& \frac{\left(1-2~\sqrt{-E_n} \right) \exp\left(4~\sqrt{-E_n}~y \right)}{\exp\left(4~\sqrt{-E_n}~y \right)-1}. \label{potjx}
\end{eqnarray}
This can be seen as either an exponential or a hyperbolic potential because the exponential on the right side of (\ref{potjx}) 
can be expressed through a hyperbolic cotangent. We have
\begin{eqnarray}
V(y) &=& \frac{1-2~\sqrt{-E_n}}{2} \left[\mbox{cotanh}\left(2~\sqrt{-E_n}~y \right)+1
\right]. \nonumber
\end{eqnarray}
A few examples of this potential are visualized in the left part of figure \ref{figj}. Since the potential depends on the 
energy, its shape will be different for each value of $n$. Next, let us focus on the solutions of our boundary-value 
problem (\ref{odejxx}), (\ref{bvpj}). These solutions can be found through the formula (\ref{solj}). Substitution of the settings 
(\ref{setj}) yields
\begin{eqnarray}
\Phi_n(y) &=& \exp\left(-\frac{3~y}{2~n}\right) \Bigg[
\exp\left(\frac{2~y}{n}\right)-1\Bigg]~ {\cal P}_{1,n}^{(1,1)}\left[\exp\left(-\frac{y}{n} \right)\right], \label{soljx}
\end{eqnarray}
where we included an index on the left side and left out irrelevant overall factors. Note that we can express these 
solutions through the definition (\ref{xjac}) of the $X_m$-Jacobi polynomials, but omit to do so for the sake of 
brevity. The right part of figure \ref{figj} visualizes solutions (\ref{soljx}) through their conventional density 
$|\Phi_n|^2$ for the first few values of $n$. 
 \noindent 
Without showing the explicit calculation let us remark that the family $(\Phi_n)$ defined in (\ref{soljx}) is orthogonal 
with respect to (\ref{ortho}). In the final step of this example we must verify that the solutions 
we constructed are physically acceptable. To this end, we will use (\ref{norm}). The integrand in the latter norm 
can be calculated by 
substituting (\ref{potjx}) and (\ref{soljx}). After evaluation of the potential's derivative and reinstalling the parameter 
$n$ through (\ref{enej}), elementary simplifications render the integrand in the form
\begin{eqnarray}
\left[1-\frac{\partial V(y)}{\partial E_n} \right] \left|\Phi_n(y) \right|^2 &=& 
16~\exp\left(-\frac{3~y}{n} \right) \Bigg\{1+\exp\left(\frac{4~y}{n}\right) +2~n+ \nonumber \\[1ex]
&+& 2~\exp\left(\frac{2~y}{n}\right) (n+1)~(2~y-1) \Bigg\}
~\left|{\cal P}_{1,n}^{(1,1)}\left[\exp\left(-\frac{y}{n} \right)\right] \right|^2 \hspace{-.2cm}. \label{denj}
\end{eqnarray}
Taking into account that the domain of our problem is the interval $(0,\infty)$, we can now calculate the norm 
(\ref{norm}) by integrating (\ref{denj}) over that domain. We arrive at
\begin{eqnarray}
\int\limits_0^\infty \left[1-\frac{\partial V(y)}{\partial E_n} \right] \left|\Phi_n(y) \right|^2 dy&=& \frac{128~n^3}{n+1},~~~n=2,4,6,... 
\label{normx}
\end{eqnarray}
Note that this result is restricted to even values of $n$. If $n$ is odd, the integral does not exist because its 
integrand is unbounded at infinity. This behaviour can be explained as follows: we see from (\ref{denj}) that the 
coefficient of the squared $X$-Jacobi polynomial produces a leading term $\sim \exp(y/n)$. For even $n$, the 
squared $X$-Jacobi polynomial does not contain any constant, such that the aforementioned leading term cancels 
out. Consequently, the integral (\ref{normx}) exists. If $n$ takes an odd value, the squared $X$-Jacobi 
polynomial contains a constant, such that the leading term $\sim \exp(y/n)$ from the coefficient is maintained. 
As a result, the integral (\ref{normx}) diverges. Finally, let us observe 
that the norms (\ref{normx}) are positive for all $n$. As an immediate consequence we obtain that our 
solutions (\ref{soljx}) are physically acceptable.

\subsection{\boldmath{$X_m$}-Laguerre polynomials} 
Our last case concerns the Laguerre case, where we for each of the three possible subcases find the Schr\"odinger form of the 
respective governing equation, together with its formal solutions. We will follow up with two examples that present 
systems admitting bound-state solutions. The second of these systems will be governed by a Dirac equation with 
pseudoscalar potential.

\subsubsection{Schr\"odinger-type equation and solutions}
Since there are three different types of 
$X_m$-Laguerre polynomials, we will work through each of these types separately before presenting our example.

\paragraph{Type I.} Our general differential equation (\ref{ode}) takes the form (\ref{odel1}) for the first type of 
$X_m$-Laguerre polynomials if the settings
\begin{eqnarray}
f(x) = -x \qquad g(x) = -1+x-\alpha+\frac{2~x~L_{m-1}^\alpha(-x)}{L_m^{\alpha-1}(-x)} \qquad 
h(x) = -n+\frac{2~\alpha~L_{m-1}^\alpha(-x)}{L_m^{\alpha-1}(-x)} \label{fl1}
\end{eqnarray}
are used. The most general Schr\"odinger form of (\ref{odel1}) that can be obtained through a point transformation 
is generated by substituting (\ref{fl1}) into (\ref{ode}). After some elementary manipulations we get
\begin{eqnarray}
\Phi''(y) + \Bigg\{ \hspace{-.4cm}
&-& \hspace{-.2cm}\frac{\left[x'(y) \right]^2}{4}+\frac{\left[x'(y) \right]^2}{4~x(y)^2}-\frac{\alpha^2 \left[x'(y) \right]^2}{4~x(y)^2}+
\frac{\left[x'(y) \right]^2}{2~x(y)}+\frac{n~\left[x'(y) \right]^2}{x(y)}+\frac{\alpha \left[x'(y) \right]^2}{2~x(y)}
- \nonumber \\[1ex]
&-& \hspace{-.2cm}
\frac{2 \left[x'(y) \right]^2 L_{m-1}^\alpha\left[-x(y) \right]^2}{L_m^{\alpha-1}\left[-x(y) \right]^2}
+\frac{\left[x'(y) \right]^2 L_{m-2}^{\alpha+1}\left[-x(y) \right]}{L_m^{\alpha-1}\left[-x(y) \right]}
-\frac{\left[x'(y) \right]^2 L_{m-1}^\alpha\left[-x(y) \right]}{L_m^{\alpha-1}\left[-x(y) \right]}
+ \nonumber \\[1ex]
&+& \hspace{-.2cm}
\frac{\left[x'(y) \right]^2 L_{m-1}^\alpha\left[-x(y) \right]}{x(y)~L_m^{\alpha-1}\left[-x(y) \right]}
-\frac{\alpha \left[x'(y) \right]^2 L_{m-1}^\alpha\left[-x(y) \right]}{x(y)~L_m^{\alpha-1}\left[-x(y) \right]}
-\frac{3 \left[x''(y) \right]^2}{4~\left[x'(y) \right]^2}+
\frac{x'''(y)}{2~x'(y)}
\Bigg\} \Phi(y) = 0. \nonumber \\ \label{odel1x}
\end{eqnarray}
In order to produce a constant term in the coefficient of $\Phi$, linear or exponential functions are a suitable 
choice, as will be shown in the example below. The formal solutions to (\ref{odel1x}) can be readily expressed 
through the point transformation (\ref{pct}), using our settings (\ref{fl1}). This yields
\begin{eqnarray}
\Phi(y) &=& \exp\left[
-\frac{1}{2}~x(y)\right] x(y)^{\frac{\alpha+1}{2}}~\frac{1}{L_m^{\alpha-1}\left[-x(y)\right]}~\sqrt{\frac{1}{x'(y)}}~
{}_1{\cal L}_{m,n}^\alpha\left[x(y) \right]. \label{soll1}
\end{eqnarray}
As before, normalizability depends critically on the choice of the coordinate change $x=x(y)$.

\paragraph{Type II.} This time we can match the two equations (\ref{ode}) and (\ref{odel2}) if we set
\begin{eqnarray}
f(x) = -x \qquad g(x) = -1+x-\alpha-\frac{2~x~L_{m-1}^{-\alpha}(x)}{L_m^{-\alpha-1}(x)} \qquad 
h(x) = 2~m-n+\frac{2~x~L_{m-1}^{-\alpha}(x)}{L_m^{-\alpha-1}(x)} \label{fl2}
\end{eqnarray}
After substituting these functions into the Schr\"odinger-type equation (\ref{odet}), it takes the form
\begin{eqnarray}
\Phi''(y) + \Bigg\{
 \hspace{-.4cm}
&-& \hspace{-.2cm}\frac{\left[x'(y) \right]^2}{4}+\frac{\left[x'(y) \right]^2}{4~x(y)^2}-\frac{\alpha^2 \left[x'(y) \right]^2}{4~x(y)^2}+
\frac{\left[x'(y) \right]^2}{2~x(y)}+\frac{(n-2~m)~\left[x'(y) \right]^2}{x(y)}+\frac{\alpha \left[x'(y) \right]^2}{2~x(y)}
- \nonumber \\[1ex]
&-& \hspace{-.2cm}
\frac{2 \left[x'(y) \right]^2 L_{m-1}^{-\alpha}\left[x(y) \right]^2}{L_m^{-\alpha-1}\left[x(y) \right]^2}
+\frac{\left[x'(y) \right]^2 L_{m-2}^{1-\alpha}\left[x(y) \right]}{L_m^{-\alpha-1}\left[x(y) \right]}
-\frac{\left[x'(y) \right]^2 L_{m-1}^{-\alpha}\left[x(y) \right]}{L_m^{-\alpha-1}\left[x(y) \right]}
- \nonumber \\[1ex]
&-& \hspace{-.2cm}
\frac{\left[x'(y) \right]^2 L_{m-1}^{-\alpha}\left[x(y) \right]}{x(y)~L_m^{-\alpha-1}\left[x(y) \right]}
-\frac{\alpha \left[x'(y) \right]^2 L_{m-1}^{-\alpha}\left[x(y) \right]}{x(y)~L_m^{-\alpha-1}\left[x(y) \right]}
-\frac{3 \left[x''(y) \right]^2}{4~\left[x'(y) \right]^2}+
\frac{x'''(y)}{2~x'(y)}
\Bigg\} \Phi(y) =0. \nonumber
\end{eqnarray}
\vspace{-.5cm}
\begin{eqnarray}
\label{odel2x}
\end{eqnarray}
Upon comparing this equation with its type I counterpart (\ref{odel1x}), we observe that the same statements regarding 
a constant term in the coefficient of $\Phi$ are valid. The solutions of the above equation are given by
\begin{eqnarray}
\Phi(y) &=&  \exp\left[
-\frac{1}{2}~x(y)\right] x(y)^{\frac{\alpha+1}{2}}~\frac{1}{L_m^{-\alpha-1}\left[x(y)\right]}~\sqrt{\frac{1}{x'(y)}}~
{}_2{\cal L}_{m,n}^\alpha\left[x(y) \right]. \label{soll2}
\end{eqnarray}
Note that we obtained this result by plugging the current settings (\ref{fl2}) into our point transformation (\ref{pct}).

\paragraph{Type III.} It remains to consider the third type of $X_m$-Laguerre polynomials. Following our 
standard procedure, we match equations (\ref{ode}) and (\ref{odel3}) by setting
\begin{eqnarray}
f(x) = -x \qquad g(x) = -1+x-\alpha+\frac{2~x~L_{m-1}^{-\alpha}(-x)}{L_m^{-\alpha-1}(-x)} \qquad 
h(x) = -n \label{fl3}
\end{eqnarray}
We obtain the Schr\"odinger equation associated with these settings by plugging them into (\ref{odet}). This gives 
\begin{eqnarray}
\Phi''(y) + \Bigg\{
 \hspace{-.4cm}
&-& \hspace{-.2cm}\frac{\left[x'(y) \right]^2}{4}+\frac{\left[x'(y) \right]^2}{4~x(y)^2}-\frac{\alpha^2 \left[x'(y) \right]^2}{4~x(y)^2}+
\frac{\left[x'(y) \right]^2}{2~x(y)}+\frac{n \left[x'(y) \right]^2}{x(y)}+\frac{\alpha \left[x'(y) \right]^2}{2~x(y)}
- \nonumber \\[1ex]
&-& \hspace{-.2cm}
\frac{2 \left[x'(y) \right]^2 L_{m-1}^{-\alpha}\left[-x(y) \right]^2}{L_m^{-\alpha-1}\left[-x(y) \right]^2}
+\frac{\left[x'(y) \right]^2 L_{m-2}^{1-\alpha}\left[-x(y) \right]}{L_m^{-\alpha-1}\left[-x(y) \right]}
-\frac{\left[x'(y) \right]^2 L_{m-1}^{-\alpha}\left[-x(y) \right]}{L_m^{-\alpha-1}\left[-x(y) \right]}
+ \nonumber \\[1ex]
&+& \hspace{-.2cm}
\frac{\left[x'(y) \right]^2 L_{m-1}^{-\alpha}\left[-x(y) \right]}{x(y)~L_m^{-\alpha-1}\left[-x(y) \right]}
+\frac{\alpha \left[x'(y) \right]^2 L_{m-1}^{-\alpha}\left[-x(y) \right]}{x(y)~L_m^{-\alpha-1}\left[-x(y) \right]}
-\frac{3 \left[x''(y) \right]^2}{4~\left[x'(y) \right]^2}+
\frac{x'''(y)}{2~x'(y)}
\Bigg\} \Phi(y) =0. \nonumber
\end{eqnarray}
As before, the solutions of this equation can be expressed as
\begin{eqnarray}
\Phi(y) &=&  \exp\left[
-\frac{1}{2}~x(y)\right] x(y)^{\frac{\alpha+1}{2}}~\frac{1}{L_m^{-\alpha-1}\left[-x(y)\right]}~\sqrt{\frac{1}{x'(y)}}~
{}_3{\cal L}_{m,n}^\alpha\left[x(y) \right], \nonumber
\end{eqnarray}
where this function was obtained by inserting the parameters settings (\ref{fl3}) into our point transformation (\ref{pct}).

\subsubsection{First example} 
\begin{figure}[h]
\begin{center}
\epsfig{file=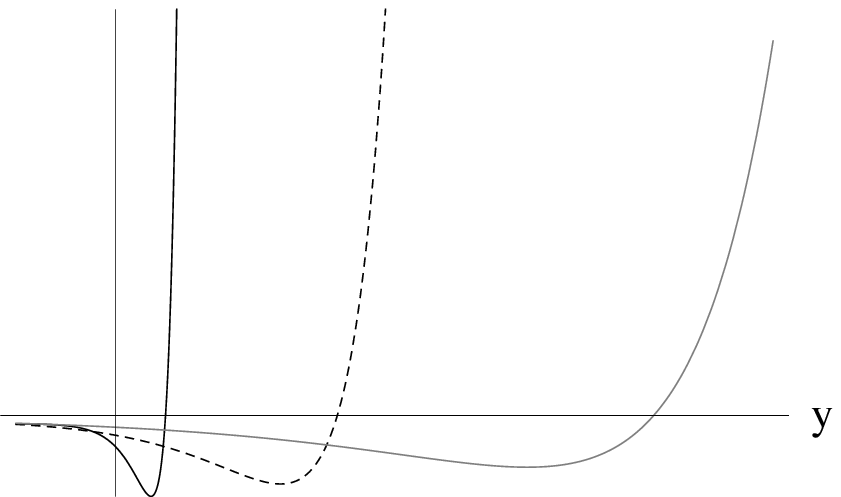,width=7.8cm}
\epsfig{file=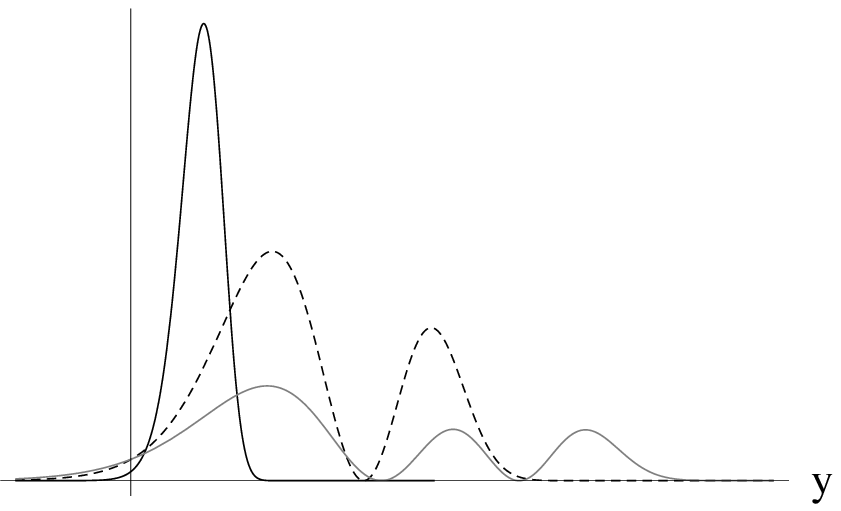,width=7.8cm}
\caption{Graphs of the potential (\ref{potl}) (left plot) and the conventional density $|\Phi_n|^2$ from 
(\ref{sollx}) (right plot) for the parameters $n=1$ (black curve), $n=2$ (dashed curve) and $n=3$ (grey curve). 
For better visibility, the plots were scaled in vertical and horizontal direction, respectively.}
\label{figl}
\end{center}
\end{figure}

We will now present a particular example involving the type I $X_m$-Laguerre polynomials. For the sake of brevity, 
examples regarding the remaining two types will not be given. We are now considering equation (\ref{odel1x}), 
where the parameters and the coordinate change need to be chosen. We set
\begin{eqnarray}
\alpha~=~5 \qquad \qquad m~=~1 \qquad \qquad x(y)~=~\exp\left(\frac{y}{n^2} \right). \label{setl}
\end{eqnarray}
Similar to previous examples, we took an exponential function as our coordinate change because it generates an 
energy term in the resulting Schr\"odinger equation. The particular dependency on $n$ renders the latter energy term 
in an especially simple form, as we will see below. Let us now substitute the settings (\ref{setl}) into 
our equation (\ref{odel1x}). We arrive at the result
\begin{eqnarray}
& &\Phi''(y)+ \Bigg\{
-\frac{29}{4~n^4}
-\frac{1}{4~n^4}~\exp\left(\frac{y}{n^2} \right) \left[
\exp\left(\frac{y}{n^2} \right)-4~(n+2)
\right]+
\frac{15}{n^4 \left[\exp\left(\frac{y}{n^2} \right)+5 \right]}- \nonumber \\[1ex]
& & \hspace{7cm} -~\frac{50}{n^4 \left[\exp\left(\frac{y}{n^2} \right)+5 \right]^2}
\Bigg\}~\Phi(y) ~=~ 0. \label{odelx}
\end{eqnarray}
This Schr\"odinger equation for an enegy-dependent potential can be assigned the domain $J=(-\infty,\infty)$ because the 
coordinate change in (\ref{setl}) maps the real axis to the positve real axis. This gives rise to the Dirichlet-type 
boundary conditions 
\begin{eqnarray}
\lim\limits_{y \rightarrow -\infty} \Phi(y) ~=~ \lim\limits_{y \rightarrow \infty} \Phi(y) ~=~ 0, \label{bvpl}
\end{eqnarray}
that we will impose on the solutions of (\ref{odelx}). We note that the coefficient of $\Phi$ in (\ref{odelx}) has a 
constant term that represents the stationary energy of the system. We identify
\begin{eqnarray}
E_n &=& -\frac{29}{4~n^4},~~~n=1,2,3,... \label{enel}
\end{eqnarray}
As in the previous examples, we included an index $n$ on the left side. Observation of (\ref{enel}) shows that 
the energies form an infinite sequence that is bounded from below by the ground state energy $E_1=-29/4$. 
In the next step, let us find the actual energy-dependent potential contained in (\ref{odelx}) by replacing $n$ as a 
function of $E_n$. This replacement is done by inversion of (\ref{enel}), that is, we have
\begin{eqnarray}
n &=& \sqrt{\frac{1}{2}} \left(-\frac{29}{E_n}\right)^\frac{1}{4}. \label{ne2}
\end{eqnarray}
Observe that the right side is real-valued because the energy $E_n$ is negative. Substitution of this expression into 
equation (\ref{odelx}) and extracting the potential gives
\begin{eqnarray}
V(y) &=& \frac{4~E_n}{29} \left[
2+\sqrt{\frac{1}{2}} \left(-\frac{29}{E_n}\right)^\frac{1}{4}\right] \exp\left(-\sqrt{-\frac{4~E_n}{29}}~y\right)
-\frac{E_n}{29}~\exp\left(-\sqrt{-\frac{16~E_n}{29}}~y\right)+ \nonumber \\[1ex]
&+&\frac{60~E_n}{29 \left[\exp\left(-\sqrt{-\frac{4~E_n}{29}}~y\right)+5\right]}
- \frac{200~E_n}{29 \left[\exp\left(-\sqrt{-\frac{4~E_n}{29}}~y\right)+5\right]^2}. \label{potl}
\end{eqnarray}
A few graphs of this potential for different values of the energy are displayed in the left part of figure \ref{figl}. 
Now, it remains to determine the solutions of the boundary-value problem (\ref{odelx}), (\ref{bvpl}). We find them by 
substituting our settings (\ref{setl}) into (\ref{soll1}). This gives 
\begin{eqnarray}
\Phi_n(y) &=& \frac{1}{\left[\exp\left(\frac{y}{n^2}\right)+5 \right]}~\exp\left[
\frac{5~y}{2~n^2}-\frac{1}{2}~\exp\left(\frac{y}{n^2}\right)\right] 
{}_1{\cal L}_{1,n}^5\left[\exp\left(\frac{y}{n^2}\right)\right], \label{sollx}
\end{eqnarray}
where again an index $n$ was included on the left side. We do not show the explicit form of the solutions in terms of 
the definition (\ref{xlag1}) because the resulting expressions would be very long. The right part of figure \ref{figl}) shows 
graphs of solutions (\ref{sollx}) by means of their conventional density $|\Phi_n|^2$ for the first few values of $n$. 
As in the previous examples, the familiy of solutions $(\Phi_n)$ defined in (\ref{sollx}) obey the orthogonality 
relation (\ref{ortho}). Now, it remains to calculate the energy-dependent norm (\ref{norm}) for the present case in order to make sure that our solutions 
are physically meaningful. To this end, we plug (\ref{setl}) and 
(\ref{sollx}) into (\ref{norm}). After evaluation of the potential's derivative and further simplification, the integrand on the 
right side of (\ref{norm}) takes the form
\begin{eqnarray}
\left[1-\frac{\partial V(y)}{\partial E_n} \right] \left|\Phi_n(y) \right|^2 &=& 
\frac{\exp\left[\frac{5~y}{n^2}-\exp\left(\frac{y}{n^2}\right)\right]}{116~n^2 \left[\exp\left(\frac{y}{n^2}\right)+5 \right]^5}~
\Bigg\{
3125~n^2+\exp\left(\frac{5~y}{n^2}\right) (n^2~y+1)+ \nonumber \\[1ex]
& & \hspace{-2.5cm}+~\exp\left(\frac{4~y}{n^2}\right) 
\Bigg[n~\Bigg(~\big(7-3~n\big)~\Bigg)~n-2~y\Bigg)+11~y\Bigg]
- \nonumber \\[1ex]
& & \hspace{-2.5cm}-~
\exp\left(\frac{3~y}{n^2}\right)
\Bigg[n^2~\Bigg(16+45~n\Bigg)+15~\Bigg(2~n-1\Bigg)~y\Bigg]
-\nonumber \\[1ex]
& & \hspace{-2.5cm}-~
25~\exp\left(\frac{y}{n^2}\right)
\Bigg[n^2~\Bigg(15~n-31\Bigg)+2~y~\Bigg(5~n+11 \Bigg)\Bigg]
-\nonumber \\[1ex]
& & \hspace{-2.5cm}-~
5~\exp\left(\frac{2~y}{n^2}\right)
\Bigg[29~y+5~n~\Bigg(n~
\big(9~n+4\big)+6~y\Bigg)\Bigg]
\Bigg\} ~\left|
{}_1{\cal L}_{1,n}^5\left[\exp\left(n^2 y\right)\right]
\right|. \label{findens}
\end{eqnarray}
In order to calculate the norm, this function must be integrated over the real line, recall that $J=(-\infty,\infty)$. 
Since we were not able to perform the 
integration symbolically, we state the numerical values of the norm for the first few values of $n$. We have
\begin{eqnarray}
\Vert \Phi_1 \Vert ~=~ 1.24137 \qquad \qquad \qquad \Vert \Phi_2 \Vert ~=~ 57.93103 \qquad \qquad \qquad 
\Vert \Phi_3 \Vert ~=~ 670.34483. \label{nums}
\end{eqnarray}
The norms keep increasing if higher values of $n$ are used, implying that they remain positive. As a result, we 
conclude that our solutions (\ref{sollx}) are physically acceptable.
 \noindent 
Since we cannot find a closed form of the norm integral over (\ref{findens}) through direct integration, let us now construct 
such a form by means of the method used in section 3.1.2. In the present case, the right side of our formula 
(\ref{nrep}) contains Wronskians of the solution $\Phi_n$ in (\ref{sollx}) and its derivative with respect to $E$. 
Recall that this derivative 
can be obtained by replacing $n$ through (\ref{ne2}). Furthermore, note that the quantities $x_0$ and $x_1$ are given 
by negative and positive infinity. In its most general form, the right side of (\ref{nrep}) takes a very complicated form, such 
that we prefer not to show it here. We will restrict ourselves to presenting the simplest case for $n=1$. This leads to 
the Wronskian
\begin{eqnarray}
W_{\Phi_n,\frac{\partial \Phi_n}{\partial E_n}}(y)_{\mid n=1} &=&
\frac{\exp\left(-\exp(y)+5~y\right)}{116~[5+\exp(y)]^4}
\Bigg\{-4500+\exp(y)~\Bigg[60~(-39+16~y)+\exp(y) \times \nonumber \\[1ex]
&\times& \Bigg(-223+660~y+\exp(y)~
\bigg\{73+179~y+\exp(y)~\bigg[17+22~y+\exp(y) \times \nonumber \\[1ex]
&\times& (1+y) \bigg] \bigg\} \Bigg) \Bigg]-
\exp(y)~\bigg[5+\exp(y) \bigg]^3 \bigg\{  \bigg[ 6+\exp(y) \bigg]~\frac{\partial}{\partial k}~L_{k}^6[\exp(y)]_{\mid k=-2}+ 
\nonumber \\[1ex]
&+& \bigg[7+\exp(y) \bigg]~
\frac{\partial}{\partial k}~L_{k}^5[\exp(y)]_{\mid k=-1})\Bigg\}.
\nonumber
\end{eqnarray}
Note that the symbols $L_k^5$ and $L_k^6$ refer to the Laguerre function \cite{abram}, defined by means of 
hypergeometric functions. As such, the Laguerre function is differentiable with respect to $k$. Now, 
application of limits at positive and negative infinity gives
\begin{eqnarray}
\lim\limits_{y \rightarrow -\infty} W_{\Phi_n,\frac{\partial \Phi_n}{\partial E_n}}(y)_{\mid n=1} ~=~0 
\qquad \qquad \qquad
\lim\limits_{y \rightarrow \infty} W_{\Phi_n,\frac{\partial \Phi_n}{\partial E_n}}(y)_{\mid n=1} ~=~ -1.24137. \nonumber
\end{eqnarray}
Substitution of these limits into (\ref{nrep}) yields
\begin{eqnarray}
\Vert \Phi_0 \Vert &=& \lim\limits_{y \rightarrow -\infty} W_{\Phi_n,\frac{\partial \Phi_n}{\partial E_n}}(y)_{\mid n=1}-
\lim\limits_{y \rightarrow \infty} W_{\Phi_n,\frac{\partial \Phi_n}{\partial E_n}}(y)_{\mid n=1} ~=~1.24137 , 
\nonumber
\end{eqnarray}
confirming the result found in (\ref{nums}).

\subsubsection{Second example: Dirac equation}
\begin{figure}[h]
\begin{center}
\epsfig{file=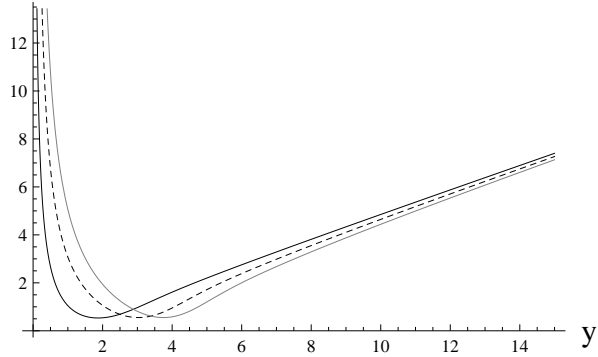,width=7.8cm}
\caption{Graphs of the parametrizing function (\ref{qc}) for $n=1$ (solid curve), $n=2$ (dashed curve) and $n=3$ 
(gray curve).}
\label{pot_dirac}
\end{center}
\end{figure}

\begin{figure}[h]
\begin{center}
\epsfig{file=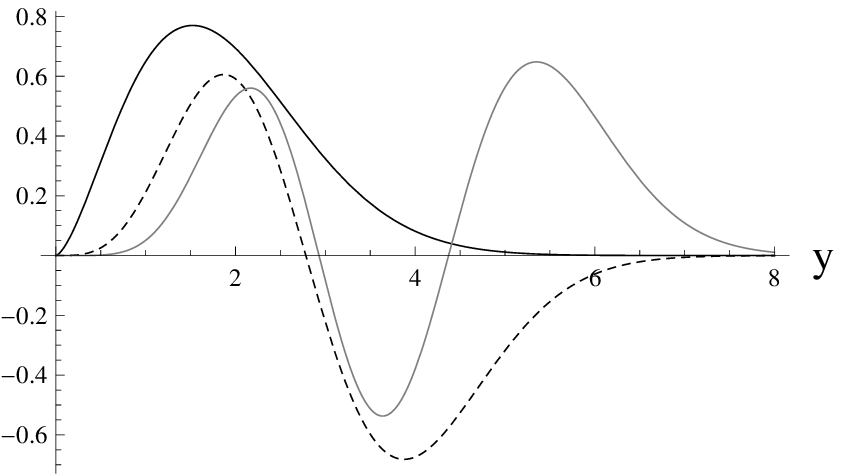,width=7.8cm}
\epsfig{file=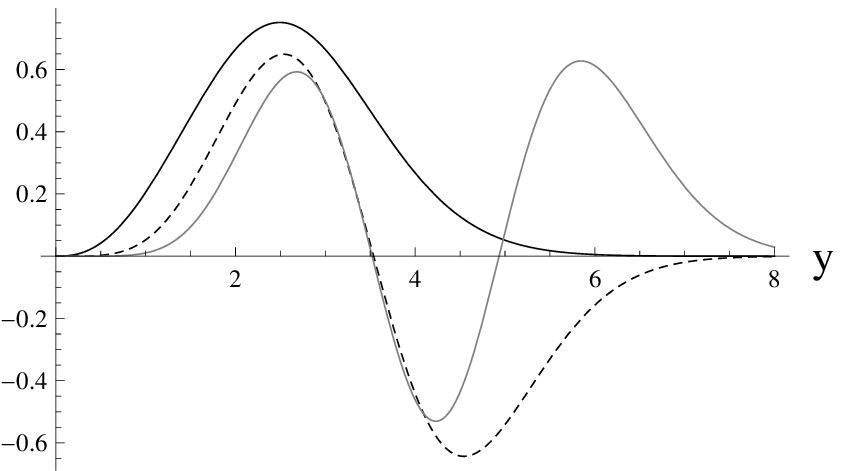,width=7.8cm}
\caption{Graphs of the first solution component (\ref{psi1nx}) (left plot) and the second solution component (\ref{psi2nx}) 
(right plot) for $n=1$ (solid curve), $n=2$ (dashed curve), $n=3$ 
(gray curve) and the overall setting $M=1$.}
\label{sol_dirac}
\end{center}
\end{figure}

In this application we will demonstrate that our method can be extended to generate solutions of the Dirac equation that are 
given in terms of type II $X_m$-Laguerre polynomials. 
To this end, let us start out from (\ref{odel2x}) by applying the following settings
\begin{eqnarray}
m~=~1 \qquad \qquad x(y)~=~\frac{y^2}{2},~~~y>0. \label{setl2}
\end{eqnarray}
This coordinate change will produce an energy term in the Schr\"odinger equation that we obtain by plugging 
(\ref{setl2}) into  (\ref{odel2x}). We get the following result
\begin{eqnarray}
& & \Phi''(y) +\Bigg\{2~n-1+\frac{1-4~\alpha^2}{4~y^2}-\frac{y^2}{4}+\frac{4}{y^2+2~\alpha}+ 
\frac{8~y^2}{\left(y^2+2~\alpha\right)^2}
\Bigg\}~\Phi(y) ~=~ 0. \label{odel2xx}
\end{eqnarray}
In accordance with the coordinate change in (\ref{setl2}), we assign the domain $J=(0,\infty)$ to equation (\ref{odel2xx}). 
Furthermore we equip it with Dirichlet-type boundary conditions 
\begin{eqnarray}
\lim\limits_{y \rightarrow 0} \Phi(y) ~=~ \lim\limits_{y \rightarrow \infty} \Phi(y) ~=~ 0. \label{bvpl2}
\end{eqnarray}
The stationary energy of the system governed by (\ref{odel2xx}), (\ref{bvpl2}) can be represented through constant terms that the 
coefficient of $\Phi$ in (\ref{odel2xx}) contains. We make the identification
\begin{eqnarray}
E_n &=& 2~n-1,~~~n=1,2,3,... \label{enel2}
\end{eqnarray}
The boundary-value problem (\ref{odel2xx}), (\ref{bvpl2}) admits a set of solutions $(\Phi_n)$ that are 
associated with the stationary energies (\ref{enel2}). These solutions are obtained by substituting our settings 
(\ref{setl2}) into equation (\ref{soll2}). We find
\begin{eqnarray}
\Phi_n(y) &=& \frac{y^{\alpha+\frac{1}{2}}}{y^2+2~\alpha}~\exp\left(-\frac{y^2}{4}\right) ~
{}_2{\cal L}_{1,n}^\alpha\left(\frac{y^2}{2}
\right). \label{phinl2}
\end{eqnarray}
We observe that the potential in (\ref{odel2xx}) is energy-dependent if and only if $\alpha$ is energy-dependent. Let us now 
transform (\ref{odel2xx}), (\ref{bvpl2}) into a boundary-value problem governed by a Dirac equation. In particular, for the 
two-component function $\Psi = (\psi_1,\psi_2)^T$ we consider the following problem
\begin{eqnarray}
i~\sigma_2~\Psi'(y)+\left[U(y)-\epsilon \right] \Psi(y) &=& 0 \label{dirac} \\[1ex]
\lim\limits_{y \rightarrow 0} \Psi(y) ~=~ \lim\limits_{y \rightarrow \infty} \Psi(y) &=& (0,0)^T, \label{diracbc}
\end{eqnarray}
where $\epsilon$ represents the stationary energy, $\sigma_j$, $j=1,2,3$, stands for the $j$-th Pauli matrix and the 
potential $U$ is assumed to be of pseudoscalar form, that is,
\begin{eqnarray}
U(y) &=& M~\sigma_3+q(y)~\sigma_1. \label{pseudo}
\end{eqnarray}
Here, the constant $M>0$ represents the mass and $q$ is the parametrizing function of the potential. It is known that 
the Dirac equation (\ref{dirac}) for the potential (\ref{pseudo}) can be transformed into a Schr\"odinger equation and 
vice versa. In the following we will use this transformation, for detailed information and derivation the reader 
is referred to e.g. \cite{xbatbar}. According to the scheme outlined in the latter reference, the function $q$ in the potential 
(\ref{pseudo}) is given by the following expression 
\begin{eqnarray}
q(y) &=& \frac{d}{dy} \left[ \Phi_{\frac{1}{2}}(y) \right]
+C \left[
c~\Phi_{\frac{1}{2}}(y)^2+\Phi_{\frac{1}{2}}(y)^2 \int\limits^y \frac{1}{\Phi_{\frac{1}{2}}(t)^2}~dt
\right]^{-1}, \label{q}
\end{eqnarray}
for two arbitrary constants $c$ and $C$. We obtain the explicit form of $\Phi_{\frac{1}{2}}$ from (\ref{phinl2}) for $n=1/2$. 
Substitution into (\ref{q}) gives 
\begin{eqnarray}
q(y) &=& \frac{y~{}_2{\cal L}_{1,\frac{1}{2}}^\alpha {'} \left(\frac{y^2}{2}\right)}{{}_2{\cal L}_{1,\frac{1}{2}}^\alpha\left(\frac{y^2}{2}
\right)}+ \frac{2 ~\alpha+1}{y}-\frac{y}{2}-\frac{2~y}{y^2+2~\alpha}+\nonumber \\[1ex]
& & \hspace{1.5cm} +~\frac{C~\exp\left(\frac{y^2}{2}\right) (y^2+2~\alpha)^2}
{y^{2 \alpha+1}~{}_2{\cal L}_{1,\frac{1}{2}}^\alpha\left(\frac{y^2}{2}
\right)^2}
\left[c+\int\limits^y 
\frac{\exp\left(\frac{t^2}{2}\right) (t^2+2~\alpha)^2}{t^{-2 \alpha-1}~{}_2{\cal L}_{1,\frac{1}{2}}^\alpha\left(\frac{t^2}{2}
\right)^2}~dt \right]^{-1} \hspace{-.3cm}. \label{qx}
\end{eqnarray}
This function determines the pseudoscalar potential (\ref{pseudo}) in our Dirac equation. 
Furthermore, the stationary energies of the Dirac equation 
(\ref{dirac}) are given by
\begin{eqnarray}
\epsilon &=& \sqrt{E_n+M^2} ~=~ \sqrt{2~n-1+M^2}, \label{de}
\end{eqnarray}
while the associated solutions $\Psi_n=(\psi_{1,n},\psi_{2,n})^T$, $n=1,2,3,...$, can be constructed by means of the relations
\begin{eqnarray}
\psi_{1,n}(y) &=& \Phi_n(y) ~=~ \frac{y^{\alpha+\frac{1}{2}}}{y^2+2~\alpha}~\exp\left(-\frac{y^2}{4}\right) ~
{}_2{\cal L}_{1,n}^\alpha\left(\frac{y^2}{2}
\right)  \label{psi1n}  \\[1ex] 
\psi_{2,n}(y) &=& \frac{1}{M+\sqrt{2~n-1+M^2}}~ \left[q(y) ~\psi_{1,n}(y)-\psi_{1,n}'(y) \right].  \label{psi2n}
\end{eqnarray}
Note that the second component $\psi_{2,n}$ of the solution can be found by substituting (\ref{qx}) and 
(\ref{psi1n}). Due to the length of the involved expressions we omit to state the explicit form of $\psi_{2,n}$. In summary, 
the functions (\ref{psi1n}), (\ref{psi2n}) form the two components of the solution to the Dirac equation (\ref{dirac}) for the 
stationary energies (\ref{de}) and the 
pseudoscalar potential (\ref{pseudo}), the parametrizing function of which is determined through (\ref{qx}). 
Before we conclude this example, let us return to the Schr\"odinger equation (\ref{odel2xx}). As mentioned before, 
its potential becomes energy-dependent if the constant $\alpha$ depends on the energy. As such, the same holds for 
the potential (\ref{pseudo}) of the Dirac equation. Let us illustrate this statement by introducing energy-dependence into 
$\alpha$ as follows
\begin{eqnarray}
\alpha(E_n) &=& E_n ~=~ 2~n-1. \label{alpha}
\end{eqnarray}
Observe that this setting is compatible with the domain of $\alpha$ that in this case is given by $\alpha > 0$. 
If we substitute (\ref{alpha}) into the Dirac equation (\ref{dirac}), its potential becomes energy-dependent. 
For the sake of simplicity we pick $C=0$ in the parametrizing function (\ref{qx}). After incorporation of (\ref{alpha}) 
we obtain
\begin{eqnarray}
q(y) &=&  \frac{y~{}_2{\cal L}_{1,\frac{1}{2}}^{2n-1} {'} \left(\frac{y^2}{2}\right)}{{}_2{\cal L}_{1,\frac{1}{2}}^{2n-1}\left(\frac{y^2}{2}
\right)}+ \frac{4~n-1}{y}-\frac{y}{2}-\frac{2~y}{y^2+4~n-2}. \label{qc}
\end{eqnarray}
Three particular cases of this potential are shown in figure \ref{pot_dirac}. These cases differ by the energy that is 
associated with them.
 \noindent 
It remains to specify the the solution components (\ref{psi1n}), (\ref{psi2n}) for the current settings. They take the form 
\begin{eqnarray}
\psi_{1,n}(y) &=& \frac{y^{2n-\frac{1}{2}}}{y^2+4~n-2}~\exp\left(-\frac{y^2}{4}\right) ~
{}_2{\cal L}_{1,n}^{2n-1}\left(\frac{y^2}{2}
\right)  \label{psi1nx}  \\[1ex] 
\psi_{2,n}(y) &=& \frac{1}{M+\sqrt{2~n-1+M^2}}~ \left[q(y) ~\psi_{1,n}(y)-\psi_{1,n}'(y) \right],  \label{psi2nx}
\end{eqnarray}
where $q$ is given in (\ref{qc}). 
These solutions are visualized in figure \ref{sol_dirac}. It is clearly visible that they satisfy the boundary conditions 
(\ref{diracbc}).
 \noindent

\section{Concluding remarks} 
We have applied a particular class of point transformations to equations solved by X-OPS. As a result, Schr\"odinger 
equations with energy-dependent potentials admitting solutions in terms of these X-OPS were generated, 
along with their formal solutions. While 
we restricted ourselves to the construction of conventional Schr\"odinger equations, a different class of 
point transformations can be used to produce more general models, for example Schr\"odinger 
equations including a position-dependent mass, solutions of which can be expressed through X-OPS. Similar to the 
present case, such models feature energy-dependent potentials. In addition, their position-dependent mass functions
generated by the point transformations will also depend on the stationary energies of the underlying system. This and related 
topics will be subject of work in progress.

\end{sloppypar}

\end{document}